\documentclass[11pt]{article}

\usepackage{epsfig}
\usepackage{amssymb}
\usepackage[]{times}
\newcommand{\be}{\begin{equation}}
\newcommand{\ee}{\end{equation}}
\newcommand{\bea}{\begin{eqnarray}}
\newcommand{\eea}{\end{eqnarray}}

\def\1#1{^{(#1)}}

\begin{document}
\title{Converting Neutron Stars into Strange Stars: Instanton Model}
\author{Victor Ts. Gurovich${}^{\dag}$ and Leonid G. Fel${}^{\ddag}$\\ \\
${}^{\dag}$Department of Physics, Technion, Haifa 32000, Israel\\
${}^{\ddag}$Department of Civil Engineering, Technion, Haifa 32000, Israel} 
\date{}
\maketitle
\def\be{\begin{equation}}
\def\ee{\end{equation}}
\def\bea{\begin{eqnarray}}
\def\eea{\end{eqnarray}}
\def\p{\prime}
\begin{abstract}
We estimate the quasiclassical probability of the homogeneous nuclear matter 
transition to a strange matter when a detonation wave propagates radially inside
a sphere of nuclear matter. For this purpose we make use of instanton method 
which is known in the quantum field theory.\\


{\bf PACS 2006:} Primary -- 47.40.Rs, 03.65.Xp, Secondary -- 26.60.-c, 97.60.Jd
\end{abstract}
\section{Introduction}\label{s1}
%%%%%%%%%%%%%%%%%%%%%%%%%%%%%%%%%%%%%%%%%%%%%%%%%%%
It was first pointed out \cite{bo71,wi84} that strange matter (SM) composed of 
three quarks might be a ground state of a normal nuclear matter (NM) at zero 
temperature and pressure, which was later supported by studies based on MIT bag 
model \cite{FJ84}. A conversion of NM to SM is suppressed at ordinary nuclear 
densities. The existence of stable SM would have some remarkable consequences in
cosmology and astrophysics. At very large densities of NM like those in neutron 
stars (NS), where the Fermi energy is higher than mass of {\em s} quark, the 
NM--SM transition may occur spontaneously. This led to conjecture \cite{al88,
bo71,ms88,wi84} that strange stars, which are predominantly made of SM, may be 
formed from dense NS. The conversion is assumed to be triggered at the core of 
NS \cite{al86,ol87} where the density reaches values $2\cdot 10^{14}g/cm^3\!<\!
\rho_*\!<\!6\cdot 10^{15}g/cm^3$ with a total mass of the star $M\geq 1.5\odot$.
There may appear stable SM drops, called {\em strangelets} \cite{FJ84}, if every
single drop possesses a baryon charge $A$ exceeding some critical value $A_*$. 
Further growth of strangelets occur by outward diffusion of strange quarks to 
ambient NM \cite{al86,ol87}.

Equation of SM state has been suggested in \cite{wi84}, $P_s=(E_s-E_o)/3$, where
$P_s$ and $E_s$ stand for the pressure and density of energy, and $E_o$ denotes 
a density of energy of SM at zero pressure. If $E_s\gg E_o$ then transition 
from the non relativistic NM ($P_n\ll E_n$) to SM occurs with essential growth 
of pressure and temperature.

There are two different models which treat the NM--SM transition in framework of
relativistic hydrodynamics: combustion waves (CW) \cite{al86, ol87} and 
detonation waves (DW) \cite{ho88, to05}. The CW propagates as a slow combustion 
with a speed $V_c\simeq 10^7m/s$, while the DW propagates with $V_d\simeq 10^8 
m/s$. In \cite{to05} DW was considered as the self-similar spherical wave 
propagating with a constant rate w.r.t. NM of constant density. Different 
aspects of this conversion were discussed in \cite{ho88, ma11}.

The problem arises when the classical solution is considered at the strangelet 
scale with radius $R_s=(3Am_n/4\pi\rho)^{1/3}$, where $\rho$ denotes a density
of NM and $m_n$ stands for neutron mass. For strangelets with baryon charge $A
\simeq 10-100$ this radius varies in the range $R_s\simeq 1.2-2.5\cdot 10^{-15}
m$. On the other hand, the de Broglie wavelength $\lambda_B=h/(Am_nV_d)$ for the
strangelet reads, $\lambda_B\simeq 0.4-4\cdot 10^{-15}m$, {\em i.e.}, both 
$\lambda_B$ and $R_s$ have comparable values. This manifests the quasiclassical 
nature of the strangelets which trigger the NM--SM transition and poses a 
question about probability of such transition. 

To answer this question we make use of the known in quantum theory instanton 
approach \cite{ra82} which describes a tunneling between different field 
configurations. We calculate a probability of the NM--SM conversion when DW 
propagates spherically inside NM.
%%%%%%%%%%%%%%%%%%%%%%%%%%%%%%%%%%%%%%%%%%%%%%%%%%%
\section{Instantons and probability}\label{s2}
%%%%%%%%%%%%%%%%%%%%%%%%%%%%%%%%%%%%%%%%%%%%%%%%%%%
An instanton is a classical non-trivial solution to equations of motion in 
${\mathbb E}^4$ with finite, non-zero action $S$. We recall the main steps of 
the instanton approach in the field theory \cite{we96}. The classical scalar 
field $\phi(x_j)$ with density $\Pi(\phi)$ of potential energy $V(\phi)$ is 
given by Lagrangian $L(\phi,x_j)=1/2\sum_{i}\left(\nabla_i\phi\right)^2-\Pi(
\phi)$. In the 4D Minkowski spacetime ${\mathbb M}^{3,1}$ the Euler-Lagrange 
equation under spherical symmetry reads
\bea
\frac{\partial^2\phi}{\partial\tau^2}-\frac1{r^2}\frac{\partial}{\partial r}
\left(r^2\frac{\partial \phi}{\partial r}\right)+\frac{\partial \Pi}{\partial
\phi}=0\;,\label{k1}
\eea
where $\tau=ct$. It has to be supplemented with boundary and initial conditions.
In the 4D Euclidean space ${\mathbb E}^4$ the time $\tau$ has to be replaced in 
(\ref{k1}) by $\vartheta=i\tau$,
\bea
\frac{\partial^2\phi}{\partial \vartheta^2}+\frac1{r^2}\frac{\partial}{\partial 
r}\left(r^2\frac{\partial \phi}{\partial r}\right)-\frac{\partial \Pi}{\partial
\phi}=0\;.\label{k2}
\eea
Then the Euclidean Lagrangian $L_e(\phi,x_j)=1/2\sum_{i}\left(\nabla_i\phi
\right)^2+\Pi(\phi)$ gives rise to the Euclidean action $S_e=\int L_e(\phi,x_j)
d^3x\;d\vartheta$. Probability $\wp$ of emergence in ${\mathbb M}^{3,1}$ of the 
non-trivial solution of equation (\ref{k1}), which is called instanton, is 
given \cite{we96} up to the pre-exponential factor,
\bea
\wp\propto\exp\left(-\frac{2|S_e|}{\hbar}\right)\;.\label{k4}
\eea
%%%%%%%%%%%%%%%%%%%%%%%%%%%%%%%%%%%%%%%%%%%%%%%%%%%%%%%%%%%%%%%%
\section{Detonation waves in relativistic hydrodynamics}\label{s3}
%%%%%%%%%%%%%%%%%%%%%%%%%%%%%%%%%%%%%%%%%%%%%%%%%%%%%%%%%%%%%%%%%%
In the case of isentropic flow, the Lagrangian $L(\phi,x_j)$ of the continuous 
matter in ${\mathbb M}^{3,1}$ can be taken \cite{st64, gu65} equal to the 
pressure $P=W-E$, where $W$ and $E$ denote the enthalpy and energy, 
respectively. Indeed, such flow allows to introduce \cite{ll87} a quasipotential
$\Phi$ such that $\Phi_{,k}=\left(W/b\right)u_k$, where $b$ denotes the density 
of baryon charge and $u_k$ is a four-velocity,
\bea
u_0=\frac1{\sqrt{1-V^2}},\quad u_j=\frac{-V_j}{\sqrt{1-V^2}},\quad j=1,2,3
,\quad V^2=\sum_{1\leq j\leq 3}V_j^2\;.\label{k5}
\eea
The Lagrangian was found in \cite{st64}, $L(\phi,x_j)=b\sqrt{\Phi_{,k}\Phi^{,k}}
-E$. Substitute into the latter the definition of $\Phi_{,k}$ and making use of 
identity $u_{,k}u^{,k}=1$ we arrive at equality $L(\phi,x_j)=P$. Such definition
is consistent \cite{gu65} with Euler equation for continuous SM,
\bea
\frac1{W_s}\left(\frac{\partial P_s}{\partial r}+V_s\frac{\partial P_s}
{\partial\tau}\right)+\frac1{1-V_s^2}\left(\frac{\partial V_s}
{\partial\tau}+V\frac{\partial V_s}{\partial r}\right)=0\;,\label{k7}
\eea
the energy conservation law,
\bea
\frac1{W_s}\left(\frac{\partial E_s}{\partial\tau}+V_s\frac{\partial E_s}
{\partial r}\right)+\frac1{1-V_s^2}\left(\frac{\partial V_s}{\partial r}+ 
V_s\frac{\partial V_s}{\partial\tau}\right)+\frac{2V_s}{r}=0\;,\label{k8}
\eea
and baryon charge conservation law,
\bea
\frac1{b_s}\left(\frac{\partial b_s}{\partial\tau}+V_s\frac{\partial b_s}
{\partial r}\right)+\frac1{1-V_s^2}\left(\frac{\partial V_s}{\partial r}+ V_s
\frac{\partial V_s}{\partial\tau}\right)+\frac{2V_s}{r}=0\;.\label{k9}
\eea

Velocity $V_s(r,\tau)$ in equations (\ref{k7}, \ref{k8}, \ref{k9}) denotes the 
radial velocity of spherical flow of SM and by $b_s$ the density of baryon 
charge in SM. These equations have to be supplemented with equation of state for
SM \cite{wi84} and for NM \cite{ll86},
\bea
P_s=\frac{E_s-E_o}{3},\quad\mbox{and}\quad P_n(\rho_n)=B\;\rho_n^{5/3}\;,\quad
B=\frac{(3\pi^2)^{2/3}}{5}\frac{\hbar^2}{m_n^{8/3}}\;,\label{k10}
\eea
Bearing in mind the similarity between equations (\ref{k7}) and (\ref{k8}) find 
the relationship between two functions $b_s(r,\tau)$ and $W_s(r,\tau)$. 
According to definition $W_s=E_s+P_s$ and formula (\ref{k10}) we get, $dE_s=3/4
\;dW_s\;.\;$ Substituting the latter into (\ref{k8}) we obtain
\bea
\frac{3}{4}\frac1{W_s}\left(\frac{\partial W_s}{\partial\tau}+V_s\frac{\partial 
W_s}{\partial r}\right)+\frac1{1-V_s^2}\left(\frac{\partial V_s}{\partial r}+
V_s\frac{\partial V_s}{\partial\tau}\right)+\frac{2V_s}{r}=0\;.\label{k11}
\eea
By comparison (\ref{k8}) and (\ref{k11}) we arrive at relationship $b_s(r,\tau)
=W_s^{3/4}(r,\tau)$.

As in the classical theory of detonation \cite{ll87}, write self-similar 
equations (\ref{k7}, \ref{k8}) assuming that the velocity $V_s(r,\tau)$ is 
depending on $r$ and $\tau$ only through the variable $\xi=r/\tau$ with 
velocity' dimension, $V_s=V_s(\xi)$. Such equations might be derived by 
replacing the differential operators in (\ref{k7}, \ref{k8}),
\bea
\frac{\partial}{\partial\tau}\to-\frac{\xi}{\tau}\;\frac{d}{d\xi},\quad
\frac{\partial}{\partial r}\to\frac1{\tau}\;\frac{d}{d\xi},\nonumber
\eea
{\em i.e.},
\bea
\left[\frac1{C_{so}^2}\left(\frac{V_s-\xi}{1-\xi V_s}\right)^2-1\right]\times
\frac{dV_s}{d\xi}=\frac{2V_s(1-V_s^2)}{\xi(1-\xi V_s)},\hspace{.6cm}
\frac1{E_s}\frac{dE_s}{d\xi}=\frac{4(\xi-V_s)}{(1-\xi V_s)(1-V^2_s)}\times
\frac{dV_s}{d\xi}\;.\label{k12}
\eea
where $C_{so}$ denotes a speed of sound in SM (in units of speed of light $c$), 
$C_{so}=\sqrt{\partial P_s/\partial E_s}=1/\sqrt{3}$. 

Both functions $V_s(\xi)$ and $E_s(\xi)$ are odd. Indeed, by replacing $V_s\to 
-V_s$, $E_s\to -E_s$, $\xi\to -\xi$ in (\ref{k12}) we arrive to the same 
equations. Boundary conditions (BC) for equations (\ref{k12}) have to be given 
at DW front where NM--SM transition occurs and the flux density of the 
energy-momentum tensor and the flux density of the baryon charge are conserved. 
According to Zeldovich' {\em normal detonation law} \cite{ll87} the DW front 
propagates w.r.t the SM with a speed of sound. 
\begin{figure}[h!]\begin{center}
\psfig{figure=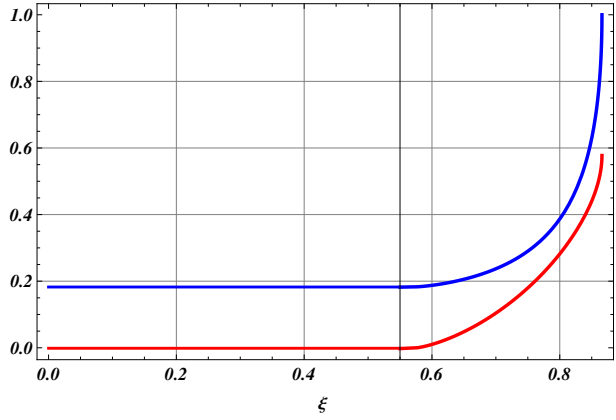,height=5.5cm}
\end{center}
\vspace{-.7cm}
\caption{Plots of the functions $V_s(\xi)$ ({\em red curve}) and $E_s(\xi)/3
E_n$ ({\em blue curve}).}\label{fg1}
\end{figure}

When the NM density reaches its value $\rho_*\simeq 10^{15}g/cm^3$ and the 
spontaneous birth of strange quarks proceeds, the NM remains yet 
non-relativistic. Indeed, in accordance with (\ref{k10}) its pressure reaches 
$P_n\simeq 6\;10^{27}J/cm^3$ that is much smaller than $E_n=\rho_*c^2\simeq 
10^{29}J/cm^3$, {\em i.e.}, $P_n\ll E_n$. According to \cite{to05} the density 
of baryon charge in NM $b_n=b_s\sqrt{6}$. Assuming that $E_s\gg E_0$, the BCs 
at the NM--SM front were derived in \cite{to05} with $\xi_0$ denoting the 
velocity of DW w.r.t. NM,
\bea
V_s(\xi_0)=1/\sqrt{3}\;,\quad E_s(\xi_0)=3E_n\;,\quad\xi_0=\sqrt{3}/2.
\label{k13}
\eea
In Figure \ref{fg1} we present the plots of the functions $V_s(\xi)$ and 
$E_s(\xi)/3E_n$ calculated numerically.
%%%%%%%%%%%%%%%%%%%%%%%%%%%%%%%%%%%%%%%%%%%%%%%%%%%
\section{Instantons in relativistic hydrodynamics}\label{s4}
\begin{figure}[h!]\begin{center}
\psfig{figure=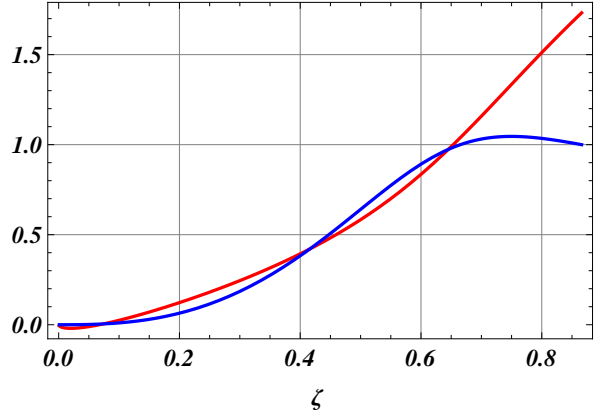,height=5.5cm}
\end{center}
\vspace{-.7cm}
\caption{Plots of the functions $\Psi(\zeta)$ ({\em red curve}) and $E_n/P_s(
\xi)$ ({\em blue curve}).}\label{fg2}
\end{figure}

%%%%%%%%%%%%%%%%%%%%%%%%%%%%%%%%%%%%%%%%%%%%%%%%%%%
We solve the Eucleadian analogue of self-similar equations (\ref{k12}) and use 
it to calculate the Eucleadian action $S_e$. Introduce a new variable, 
$\vartheta=i\tau$ which lead to new self-similar variable, $\zeta=r/\vartheta
=-i\xi$ and new velocity function $U_s=dr/d\vartheta=-iV_s$. Substitute it 
into (\ref{k12}) and get
\bea
\left[1+\frac1{C_s^2}\left(\frac{U_s-\zeta}{1+\zeta U_s}\right)^2\right]\times
\frac{dU_s}{d\zeta}=-\frac{2U_s\left(1+U_s^2\right)}{\zeta(1+\zeta U_s)}\;,\quad
\frac{d\;\ln P_s}{d\zeta}=\frac{4}{1+U_s^2}\frac{U_s-\zeta}{1+\zeta U_s}\times
\frac{dU_s}{d\zeta}\;,\label{k14}
\eea
where $E_s$ was replaced in (\ref{k12}) by $3P_s$ since $E_s\gg E_o$. Making 
use of oddness property write BCs for equations (\ref{k14}) as follows, $U_s(
\xi_0)=1/\sqrt{3}$, $P_s(\xi_0)=E_n$. The function $U_s(\zeta)$ has a singular 
point $\zeta=0$ and therefore it is convenient to introduce its inverse 
$\Psi(\zeta)=1/U_s(\zeta)$ satisfying equation,
\bea
\frac{d\Psi}{d\zeta}=\frac{2}{\zeta}\frac{\left(1+\Psi^2\right)(\zeta+\Psi)}{3
(1-\zeta\Psi)^2+(\zeta+\Psi)^2},\quad\Psi(\xi_0)=\sqrt{3}\;.\label{k15}
\eea
Equation (\ref{k15}) has no analytical solution in the range $[0,\sqrt{3}/2]$, 
however it can be found for $\zeta\ll 1$, where $\zeta, \Psi\to 0$. It reads, 
$\Psi(\zeta)=C_1\zeta^{2/3}+{\cal O}(\zeta)$, $C_1>0$. Substituting $\Psi(\zeta
)$ into (\ref{k14}) we obtain
\bea
\frac1{P_s}\frac{dP_s}{d\zeta}=-\frac{8}{\zeta}\frac{1-\zeta\Psi}{3(1-\zeta\Psi)
^2+(\zeta+\Psi)^2},\quad P_s(\xi_0)=E_n\;.\label{k16}
\eea
Numerical solutions of equations (\ref{k15}, \ref{k16}) are presented at Figure 
\ref{fg2}. The function $P_s(\zeta)$ determines the distribution of pressure in 
${\mathbb E}^4$ and allows to calculate the Euclidean action $S_e$ 
which enters into (\ref{k4}).

To show this we prove a coincidence of the Lagrangian $L(\phi,x_j)$ with the 
pressure $P_s$. Indeed, extend the Lagrangian $L(\phi,x_j)$ analytically over 
complex time $\tau=\vartheta/i$. Substituting $V_s=iU_s$ into (\ref{k5}) we find
that $u_{,k}u^{,k}=1$ in ${\mathbb E}^4$ as well as in ${\mathbb M}^{3,1}$ 
space. Thus, we obtain $L_e(\phi,x_j)=P_s$.

Start with Lorentzian action $S$ for self-similar DW. Make use of equality 
$L(\phi,x_j)=P$ and find
\bea
S=\frac{4\pi}{c}\int_0^{cT_1}\int_0^{R(T_1)}P_s\left(\frac{r}{\tau}\right)r^2
drd\tau=\frac{4\pi}{c}\int_0^{\xi_0}P_s(\xi)\xi^2d\xi\int_0^{cT_1}\tau^3d\tau.
\label{k18}
\eea
The radius $R(T_1)=\xi_0cT_1$ of SM sphere determines the DW front which 
propagates toward NM with velocity $\xi_0c$ during a time $T_1$ in such a way 
that a pressure vanishes at the front $P_s(r)=0$, $r>R$. The r.h.s. in 
(\ref{k18}) is written by rescaling to the self-similar variable $\xi$.

Write the Eucleadian action $S_e$ by replacing $\tau\to\vartheta/i$ and $P_s(
\xi)\to P_s(\zeta)$ and normalizing $P_s(\zeta)=P_s(\zeta_0)p(\zeta)$ where 
$P_s(\zeta_0)=P_s(\xi_0)$ and $p(\zeta_0)=1$. Substitute the last into 
(\ref{k18}) and integrate numerically,
\bea
S_e=-i\pi c^3P_s(\xi_0)T_1^4J(\xi_0),\quad J(\xi_0)=\int_0^{\xi_0}p(\zeta)
\zeta^2d\xi\simeq 0.6514.\label{k19}
\eea
Estimate $S_e$ by following consideration. In NS with the total mass $M\geq 
1.5\odot$ and density $\rho_*\simeq 10^{15}g/cm^3$ the spontaneous conversion 
of NM to SM is expected when a density of baryon charge reaches $n_A\simeq 6
\cdot 10^{38}cm^{-3}$. Since the spherical DW front propagates with velocity 
$\xi_0c$ then the total baryon charge grows as $A=4\pi/3\;n_A(\xi_0cT_1)^3$, 
i.e.,
\bea
T_1=\frac1{\xi_0c}\sqrt[3]{3A/(4\pi n_A)}.\label{k20}
\eea
Keep in mind an equality $P_s(\xi_0)\simeq 3m_nn_Ac^2$ which follows from 
(\ref{k13}), and combine it with (\ref{k19}),
\bea
\frac{|S_e|}{\hbar}=\frac{4r_BA^{4/3}}{3\lambda_c}J(\xi_0).\label{k21}
\eea
where $m_n$ denotes a mass of neutron, $\lambda_c=\hbar/m_nc\simeq 2.1\cdot 10^{
-14}cm$ stands for the Compton wavelength and $r_B\simeq\sqrt[3]{3/(4\pi n_A)}$ 
denotes an average distance between neutrons in NS core (for $n_A\simeq 6\cdot 
10^{38}cm^{-3}$ we have $r_A\simeq 7.4\cdot 10^{-14}cm$). 

In the WKB approximation a frequency ${\cal V}$ of emergence of the DW during 
the spontaneous conversion of $A$ neutrons to SM reads, ${\cal V}=\wp/T_1$ where
the barrirer transparency $\wp$ is given in (\ref{k4}). Appearance of the 
strangelets in NS core gives rise to propagate of DW and leads to explosion of 
NS. This scenario is realized during the time existence $T_2$ of NS and allows 
to estimate the necessary value of $A$. According to astrophysical observations 
\cite{ya99} the largest time existence of NS is approximately $10^6$ years but 
not exceeding the universe age $13.8\cdot 10^9$ years, i.e., we have $3.15\cdot 
10^{13}s<T_2<4.35\cdot 10^{17}s$. The entire number of {\em potential} 
strangelets in the NS core is given by $N_A/A$ where $N_A=M_c/m_n$ denotes a 
number of neutrons in the NS core and the mass $M_c$ of core is estimated as 
$1\%$ of the total NS mass, $M_c\simeq 10^{-2}\cdot 1.5\odot$ \cite{ya99}. Then 
a probability ${\mathbb P}$ to have at least one strangelet in core during the 
time $T_2$ is dependent on $A$ and reads
\bea
{\mathbb P}(A)=\frac{N_A}{A}\frac{T_2}{T_1}\exp\left(-\frac{2|S_e|}{\hbar}
\right),\quad S_e=S_e(A),\quad T_1=T_1(A),\quad T_2=10^{t(A)}.\label{k22}
\eea

To find a lower and upper bounds for critical value $A_*$ providing an 
appearance at least one strangelet in the core let us require ${\mathbb P}(A_*)
=1$. Solving this transcendental equation for the lower and upper bounds of 
$T_2$ we get: $23.8<A_*<24.61$. These values are pretty close to $A=20$ used in 
\cite{he96,ms95} for calculation of the ground state of strangelets in the 
framework of the MIT bag model and $A=16$ taken from space-based particle 
physics experiments on the Alpha Magnetic Spectrometer \cite{ch03} during the 
Space Shuttle Discovery mission in 1998.
%%%%%%%%%%%%%%%%%%%%%%%%%%%%%%%%%%%%%%%%%%%%%%%%%%%
\section{Concluding Remarks}\label{s5}
In the framework of instanton approach we have shown that NS with the core 
density $\rho_*\simeq 10^{15}g/cm^3$ allows to have at least one stable 
strangelet during the time star existence $T_2$, $T_N<T_2<T_U$, if the baryon 
number is $A_*=24$, where $T_N\simeq 10^6$ years and $T_U\simeq 13.8\cdot 10^9$ 
years stand for the largest time NS existence and the universe age, 
respectively. A low value of $A_*$ makes it interesting to compare it with 
those discussed in literature. 

For $2\!<\!A\!<\!6$ quantum chromodynamics strongly suggests complete 
instability of any strangelets \cite{jw99}. In \cite{FJ84} the SM is studied for
low $A<\!10^2$ and large $10^2\!<\!A\!<\!10^7$ baryon numbers. This wide range 
covers many other values for $A$ discussed in literature: $A\simeq 16-40$ 
\cite{bo71}, $A<10^2$ \cite{gj93}, $A>10^2$ \cite{ol87}, $A\simeq 10^3$ 
\cite{wi84}, $A\simeq 10^2-10^4$ \cite{ms02} and most of these values are 
substantially larger than $A_*$. We put forward an agent which may be 
responsible for the higher $A_*$ in the framework of NM--SM instanton 
transition.

\begin{figure}[h!]\begin{center}
\psfig{figure=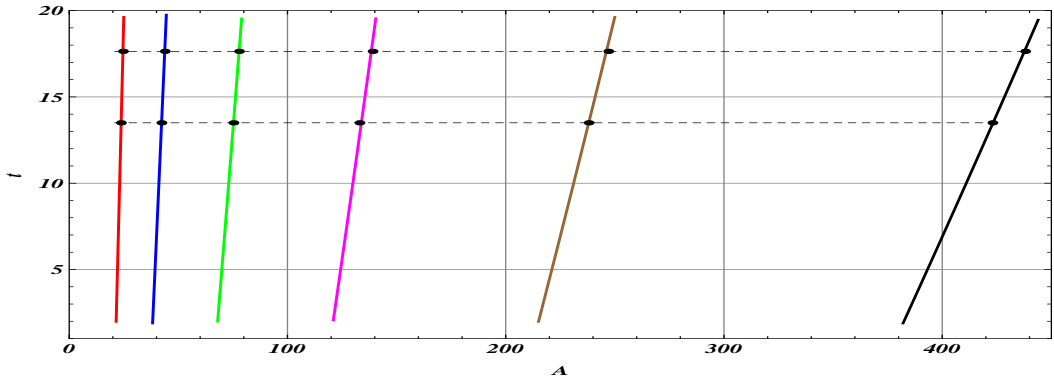,height=5cm,width=14cm}
\end{center}
\vspace{-.7cm}
\caption{Plots of the functions $t(A)$ defined in (\ref{k22}) for different 
densities $\rho_*$ ($g/cm^3$) in the NS core: $10^{15}$ ({\em red}), $10^{16}$ 
({\em blue}), $10^{17}$ ({\em green}), $10^{18}$ ({\em magenta}), $10^{19}$ 
({\em brown}), $10^{20}$ ({\em black}). Two {\em dashed lines} mark two time 
scales $T_N=10^{t_n}$, $t_n\simeq 13.5$, and $T_U=10^{t_u}$, $t_u\simeq 17.64$, 
and intersect the lines at {\em black points}.}\label{fg3}
\end{figure}

The mass of equilibrium configuration of cold matter at each central star 
density $\rho_*$ ($g/cm^3$) is a damped periodic function of $\ln\rho_*$ 
\cite{ha65,ze96}. There are two ranges for which these configurations are stable
: the white dwarfs with low electron density, $10^5<\!\rho_*<\!10^8$, and the 
neutron stars with high density, $10^{14}\!<\!\rho_*\!<\!\rho^{OV}$ where 
$\rho^{OV}\!\simeq\! 6\cdot 10^{15}$ denotes the Oppenheimer-Volkoff limit. 
There are also a number of extrema for $\rho_*$ exceeding $\rho^{OV}$: such 
superdense configurations were found in \cite{mz64}, $10^{18}\!<\!\rho_*\!<\!
10^{20}$, and in \cite{ha65}, $\rho_*>3\cdot 10^{21}$. In Figure \ref{fg3} we 
show how the admissible values of $A$ do increase once the density of the NS 
core grows, e.g., for $\rho_*\simeq 10^{19}$ we have $240<A_*<250$.

In fact, all superdense configurations with $\rho_*>\rho^{OV}$ are metastable
due to the acoustic vibrations \cite{dk63, ha65} propagating in stars with 
characteristic time $T_a=(\gamma\bar{\rho})^{-1/2}$ where $\gamma$ denotes a 
gravitational constant and $\bar{\rho}=M_{NS}/V_{NS}$ denotes an average density
of NS of the total mass $M_{NS}$ and volume $V_{NS}$. E.g., if $\rho_*\simeq 
10^{19}$ then according to \cite{ha65} $\bar{\rho}\simeq 0.25\cdot 10^{15}$ and 
finally we have $T_a\simeq 2\cdot 10^{-4}$s. Simple calculation by formula 
(\ref{k22}) with $T_2=T_a$ and ${\mathbb P}(A_*)=1$, gives a value of $A_*$ that
provides to have at least one strangelet in core during the time $T_a$, i.e., 
$A_*\simeq 200$. Suggestions of superdense stars with core density above $\rho
^{OV}$ continue to appear in the literature \cite{gl97, pr03}.
%%%%%%%%%%%%%%%%%%%%%%%%%%%%%%%%%%%%%%%%%%%%%%%%%%%
\section*{Acknowledgement}
We appreciate useful discussions with A.M. Polyakov. The research was supported 
in part (LGF) by the Kamea Fellowship.
%%%%%%%%%%%%%%%%%%%%%%%%%%%%%%%%%%%%%%%%%%%%%%%%%%%

\end{document}